\documentclass[aps, pra, amsfonts, amssymb, amsmath, twocolumn]{revtex4-1}

\pdfoutput=1
\usepackage{amsfonts}
\usepackage{amssymb}
\usepackage{amsmath}
\usepackage{hyperref}           
\hypersetup{colorlinks = True, 
            allcolors = {blue}
}
\usepackage{natbib}
\usepackage{relsize}            
\usepackage{graphicx}                  
\usepackage[caption=false]{subfig}     

\newcommand{\bra}[1]{ \langle{#1}| }
\newcommand{\ket}[1]{ |{#1}\rangle }

\def\rVac{ |\text{vac} \rangle }

\def\rAGP{|\text{AGP}\rangle}
\def\lAGP{\langle \text{AGP} |}

\newcommand{\Cdag}[1]{ {c}^{\dagger}_{#1} }
\newcommand{\Cp}[1]{ {c}_{#1} }

\def\GamD{ \mathlarger{\mathbf{{\Gamma}}}^{\dagger} }
\def\Gam{ \mathlarger{\mathbf{{\Gamma}}} }
\newcommand{\Pdag}[1]{\mathbf{P}^{\dagger}_{#1} }
\newcommand{\N}[1]{ \mathbf{N}_{#1} }
\newcommand{\Pp}[1]{ \mathbf{P}_{#1} }
\newcommand{\Killer}[1]{\mathbf{K}_{#1}}
\newcommand{\KillerAd}[1]{\mathbf{K}^{\dagger}_{#1}}

\newcommand{\Jst}[1]{\boldsymbol{J}_{#1} }
\newcommand{\PHoppers}{\boldsymbol{\mathcal{T}}}
\newcommand{\opA}{\boldsymbol{A}}

\newcommand{\E}[1]{\langle {#1} \rangle}
\newcommand{\Eagp}[1]{\lAGP {#1} \rAGP}

\newcommand{\bigO}[1]{$\mathcal{O}(#1)$}

\newcommand{\Ham}{\boldsymbol{H}} 
\newcommand{\simHam}{\bar{\Ham}}
\newcommand{\Commute}[2]{[{#1},{#2}]}
\newcommand{\norm}[1]{\left\lVert#1\right\rVert}

\newcommand{\Eq}[1]{Eq.~({#1})}
\newcommand{\Sec}[1]{Sec.~{#1}}
\newcommand{\Fig}[1]{Fig.~{#1}}
\newcommand{\Reference}[1]{Ref.~{#1}}
\newcommand{\Apdx}[1]{Appendix.~{#1}}

\begin{document}

\title{Exploring non-linear correlators on AGP}
\author{Armin Khamoshi}
    \email[Correspondence email address: ]{armin.khamoshi@rice.edu}
    \affiliation{Department of Physics and Astronomy, Rice University, Houston, TX 77005-1892}

\author{Guo P. Chen}
    \affiliation{Department of Chemistry, Rice University, Houston, TX 77005-1892}
    
\author{Thomas M. Henderson}
    \affiliation{Department of Chemistry, Rice University, Houston, TX 77005-1892}
    \affiliation{Department of Physics and Astronomy, Rice University, Houston, TX 77005-1892}
    
\author{Gustavo E. Scuseria}
    \affiliation{Department of Chemistry, Rice University, Houston, TX 77005-1892}
    \affiliation{Department of Physics and Astronomy, Rice University, Houston, TX 77005-1892}

\date{\today} 

\begin{abstract}
Single-reference methods such as Hartree-Fock-based coupled cluster theory are well known for their accuracy and efficiency for weakly correlated systems. For strongly correlated systems, more sophisticated methods are needed. Recent studies have revealed the potential of the antisymmetrized geminal power (AGP) as an excellent initial reference for the strong correlation problem. While these studies improved on AGP by linear correlators, we explore some non-linear exponential ans\"atze in this paper. We investigate two approaches in particular. Similar to \textit{Phys. Rev. B} \textbf{91}, 041114(R) (2015), we show that the similarity transformed  Hamiltonian with a Hilbert-space Jastrow operator is summable to all orders and can be solved over AGP by projecting Schr\"odinger's equation. The second approach is based on approximating the unitary pair-hopper ansatz recently proposed for application on a quantum computer. We report benchmark numerical calculations against the ground state of the pairing Hamiltonian for both of these approaches. 
\end{abstract} 


\maketitle

\section{Introduction} \label{sec:introduction}
Many-body methods in electronic structure theory frequently start from a mean-field reference. Commonly, a single-reference Slater determinant obtained by Hartree-Fock (HF) calculations is used for post-HF methods that make correction to this mean-field \cite{helgaker_molecular_2000}. Such methods include configuration interaction (CI) and coupled cluster theory (CC). For weakly correlated systems, single-reference CC is often regarded as the \textit{gold standard} and calculations are routinely performed using CC with single and double excitations (CCSD) very accurately at an affordable cost of \bigO{M^6} where $M$ is the size of the system \cite{scuseria_efficient_1988, bartlett_coupled-cluster_2007, helgaker_molecular_2000}. Unfortunately, the same cannot be said for the strong correlation problem. When many-body interactions are strong, more than one Slater determinant becomes important and methods based on a single HF reference often break down \cite{bulik_can_2015, degroote_polynomial_2016, qiu_projected_2017}. Although multireference methods exist, they have certain limitations and are not always straightforward to use \cite{roos_complete_1980, olsen_casscf_2011, szalay_multiconfiguration_2012, lyakh_multireference_2012}. As such, a new a way of thinking and more sophisticated methods are desirable.

A useful way of organizing Hilbert space in the context of strong correlation is by \textit{seniority} \cite{ring_nuclear_1980, bytautas_seniority_2011}. Loosely speaking, seniority ($\Omega$) is the number of unpaired fermions in some appropriate pairing scheme.  For example, a two-body Hamiltonian can generally be written as \cite{henderson_pair_2015}
\begin{align}
    \Ham = \Ham^{\delta \Omega= 0} + \Ham^{\delta \Omega= 2} + \Ham^{\delta \Omega=4}    
\end{align}
where the superscripts indicate the portion of the Hamiltonian that couples those determinants whose seniority differs by $0, 2$, etc. Numerical evidence has shown that the low-seniority portions of the Hilbert space can recover a very large part of the correlation energy and have the correct qualitative behavior \cite{bytautas_seniority_2011, bytautas_seniority_2015}. The seniority-zero subspace can be solved exactly by doubly occupied configuration interaction (DOCI) but it scales combinatorially with system size \cite{veillard_complete_1967, couty_generalized_1997, kollmar_new_2003}. Remarkably, CC restricted to paired double excitation (pCCD) \cite{stein_seniority_2014}, also known as AP1roG \cite{limacher_new_2013}, is often nearly exact in this subspace with a very low computational cost for realistic molecular Hamiltonians \cite{johnson_size-consistent_2013, henderson_seniority-based_2014, shepherd_using_2016}. However, this correspondence is not universal and it has been shown that pCCD and its extensions break down for some problems including the attractive pairing Hamiltonian \cite{dukelsky_colloquium:_2004, henderson_quasiparticle_2014, henderson_pair_2015, degroote_polynomial_2016, henderson_correlating_2020},
\begin{align}\label{eq:bcs_Hamiltonian}
    \Ham = \sum_{p} \epsilon_p \N{p} - G \sum_{pq} \Pdag{p}\Pp{p},
\end{align}
where $\epsilon_p = p \Delta \epsilon$ are equally spaced, doubly degenerate energy levels where $\Delta \epsilon$ is the level spacing, $G$ tunes the strength of the infinite-range pairwise interactions, and 
\begin{subequations} \label{eqs:generators}
    \begin{align}
        \Pdag{p} &= \Cdag{p}\Cdag{\bar{p}}, \\
        \N{p}    &= \Cdag{p}\Cp{p} + \Cdag{\bar{p}}\Cp{\bar{p}},
    \end{align}
\end{subequations}
such that $\Cdag{p}$ creates a fermion in spin-orbital $p$, and $\bar{p}$ is its ``paired" companion. Generally, in the seniority-zero subspace, all operators can be organized as a linear combination of terms like $\Pdag{p}\ldots\N{q}\ldots\Pp{r}$ \cite{khamoshi_efficient_2019}, where the \textit{nilpotent} operators $\Pdag{p}$ and $\Pp{p}$ together with $\N{p}$ are generators of a $su(2)$ Lie algebra 
\begin{subequations}\label{eq:CommutationRelations}
\begin{align}
\left[\Pp{p}, \Pdag{q}\right] &= \delta _{pq}\left( 1- \N{p}\right), \\
\left[\N{p}, \Pp{q}^{\dagger}\right] &= 2\delta _{pq} \Pdag{q}.
\end{align}
\end{subequations}

The breakdown of CC in the pairing Hamiltonian can be understood in terms of the spontaneous breaking of number symmetry in the electronic mean-field, which occurs at some critical value of $G$, $G_c>0$. As a result, the mean-field solution gives rise to the number-broken Bardeen–Cooper–Schrieffer (BCS) wavefunction \cite{bardeen_theory_1957} for $G>G_c$ and a number-conserving Slater determinant for $G<G_c$ including all $G<0$ where the interaction is repulsive. Indeed, it is in the attractive regime where CC as well as a whole host of other conventional methods struggle \cite{henderson_correlating_2020}. It is noteworthy that the pairing Hamiltonian is exactly solvable \cite{richardson_exact_1964,dukelsky_colloquium:_2004}.

Recently, we proposed to use the AGP wavefunction as the initial reference for this and potentially more general problems \cite{henderson_geminal-based_2019, khamoshi_efficient_2019, henderson_correlating_2020, harsha_wave_2020, dutta_geminal_2020,khamoshi_correlating_2020}. Along similar lines of research, Johnson et al. have developed correlated methods by explicitly using the pairing Hamiltonian's eigenvectors, Richardson-Gaudin states, as the initial reference \cite{johnson_richardsongaudin_2020,fecteau_reduced_2020, fecteau_richardson-gaudin_2020, johnson_transition_2020}. The main advantage of AGP for the seniority-zero sector is that on the one hand, it puts coefficients on each Slater determinant in the DOCI while still conserving seniority, and on the other hand, AGP is computationally straightforward. To put it differently, the natural orbitals of AGP, HF, pCCD, and DOCI are all the same if the same pairing scheme is chosen, differing only in occupation. The AGP wavefunction can then be expanded in its seniority-zero natural orbital determinants and can occupy all $M \choose N$ such determinants, where $N$ is the number of fermion pairs. It thus encompasses all possible configurations in the seniority-zero space, albeit with inexact coefficients. However, unlike DOCI, AGP can be optimized at mean-field cost \bigO{M^3} \cite{scuseria_projected_2011}. It also encompasses the HF state, so it has a much richer structure as an initial reference. Moreover, we have shown recently that the reduced density matrices (RDM) over AGP can be computed efficiently as all higher order RDMs are decomposable in terms of lower order ones. \cite{khamoshi_efficient_2019} These realizations permit us to envision \textit{post}-AGP methods---analogous to post-HF---to capture the remaining error and provide accurate ans\"atze in both weak and strong correlation. A case can be made for using AGP in the seniority non-zero spaces where pairs are broken, but we concentrate on the seniority-zero case at this stage.

Recent studies in our group have developed CI-based methods over AGP and numerical calculations show excellent accuracy when applied to the pairing Hamiltonian. These methods include orthogonal excitations \cite{henderson_geminal-based_2019, henderson_correlating_2020} and those based on the generators of the Lie algebra \Eq{\ref{eq:CommutationRelations}} \cite{dutta_geminal_2020}. It has been shown that these models are formally equivalent and can be viewed as ``geminal replacement" methods \cite{dutta_geminal_2020}. Encouraged by the accuracy of linear models, we would like to explore non-linear correlator ans\"atze over AGP. Non-linear models in principle can approximate the DOCI coefficient differently and it remains to be seen whether they provide an advantage over the linear case. Developing a CC theory over AGP is not trivial and remains an open problem \cite{henderson_correlating_2020}.

In this paper, we explore two exponential correlators over AGP which we build from number-conserving combinations of the generators of the Lie algebra, \Eq{\ref{eq:CommutationRelations}}. The first is the exponential of Jastrow \cite{jastrow_many-body_1955} operators on AGP (JAGP). Jastrow-Gutzwiller correlators are ubiquitous in physics and chemistry and are widely used in Monte Carlo calculations. \cite{jastrow_many-body_1955, haldane_exact_1988, foulkes_quantum_2001, casula_geminal_2003, casula_correlated_2004, umezawa_transcorrelated_2003, tsuneyuki_transcorrelated_2008, henderson_linearized_2013, neuscamman_communication_2013, neuscamman_improved_2016, genovese_assessing_2019,nakano_all-electron_2019} It was shown in \Reference{\cite{wahlen-strothman_lie_2015, wahlen-strothman_biorthogonal_2016}} that the similarity transformation of any fermionic Hamiltonian, $\mathrm{e}^{-\Jst{}} \Ham \mathrm{e}^{\Jst{}}$, where $\Jst{}$ is a Hilbert-space Jastrow operator, is analytically summable to all orders. Inspired by that, we extend the formalism to the pairing $su(2)$ algebra \Eq{\ref{eq:CommutationRelations}} and show that we can efficiently solve the projected Schr\"odinger equation over AGP. In a way, this could be viewed as a form of \textit{generalized} coupled cluster theory, but instead of the traditional particle-hole excitations, correlation is mediated by number operators that create excitations on AGP---a possibility that does not occur for single Slater determinants as they are eigenfunctions of $\Jst{}$.

The second ansatz is the unitary \textit{pair-hopper} that we proposed recently for applications in quantum computers \cite{khamoshi_correlating_2020}. Unitary ans\"atze can be implemented efficiently on a quantum computer, \cite{cao_quantum_2019} but some approximations must be made in order to implement them on a classical computer. Here, we approximate the unitary transformation of the Hamiltonian using the Baker-Campbell-Hausdorff (BCH) formula by truncating it at the 4th commutator. Affordable computational scaling is made possible by the \textit{reconstruction formulae} \cite{khamoshi_efficient_2019} with which we can eliminate the bottleneck of computing and storing higher rank density matrices from our calculations. As an alternative to the brute-force expansion of the BCH formula, we apply the \textit{canonical transformation} theory as formulated in \Reference{\cite{yanai_canonical_2006, yanai_canonical_2007, neuscamman_quadratic_2009, neuscamman_review_2010, yanai_extended_2012}} to recursively sum the transformed Hamiltonian. We report benchmark calculations of all methods against the ground state energy of the pairing Hamiltonian.

The organization of this paper is as follows. To familiarize the reader with our notation, in \Sec{\ref{sec:background}} we review some basic properties of AGP and its density matrices. In \Sec{\ref{sec:JAGP}}, we present our formalism for solving similarity transformed JAGP as well as its application to the pairing Hamiltonian. \Sec{\ref{sec:unitaryPH}} discusses the truncated commutator expansion and canonical transformation with the unitary pair-hoppers ansatz over the pairing Hamiltonian. Lastly, some discussion and concluding remarks are provided in \Sec{\ref{sec:conclusions}}. 

\section{Background} \label{sec:background}
The AGP wavefunction with $N$ pairs corresponds to a condensate where all $2N$ fermions are placed in the same geminal. Mathematically it can be written as \cite{coleman_structure_1965, surjan_introduction_1999}
\begin{align} \label{eq:AGP_as_geminal}
    \rAGP = \frac{1}{N!} \left( \GamD \right)^N \rVac,
\end{align}
where $\rVac$ is the physical vacuum and $\GamD$ is the geminal creation operator whose general expression is
\begin{align}
    \GamD = \sum_{pq} \eta_{pq} \Cdag{p} \Cdag{q}.
\end{align}
Here, $p,q$ denote spin-orbitals and $\eta$ is a skew-symmetric matrix known as the \textit{geminal coefficient}. To make a connection with \Eq{\ref{eqs:generators}} and for mathematical convenience, we perform an orbital rotation that brings the matrix of geminal coefficients into a block-diagonal form \cite{hua_theory_1944}, that is
\begin{align}
    \eta = \bigoplus_{p=1}^{M}     
    \begin{pmatrix}
        0 & \eta_p \\
        -\eta_p & 0 \\ 
    \end{pmatrix}.
\end{align}
Therefore, without loss of generality, we obtain the geminal operator in the \textit{natural orbital} basis of the geminal as
\begin{align}
    \GamD = \sum_{p = 1}^{M} \eta_{p} \Cdag{p} \Cdag{\bar{p}} = \sum_{p = 1}^{M} \eta_{p} \Pdag{p}.
\end{align} 
where we made use of the notation in \Eq{\ref{eqs:generators}} and $M$ denotes the number of orbitals. In this formalism, we can express the AGP wavefunction as 
\begin{align}\label{eq:AGP_expanded}
    \rAGP = \sum_{ 1 \leq p_1<...<p_N \leq M} \eta _{p_1}...\eta _{p_N} \Pdag{p_1}...\Pdag{p_{N}} \rVac.
\end{align}
Similarly, we can write down the equation for DOCI,
\begin{align}\label{eq:DOCI}
    \ket{\text{DOCI}} = \sum_{ 1 \leq p_1<...<p_N \leq M}  D_{p_1...p_N}\Pdag{p_1}...\Pdag{p_{N}} \rVac.
\end{align}
From these expressions, it is evident that the two wavefunctions have the same structure but differ in the coefficients of each determinant; in DOCI the amplitude tensor is fully connected, whereas in AGP it is fully factorized. 

Despite being a linear combination of a combinatorial number of determinants, reduced density matrices of many-body fermionic operators over AGP can be computed efficiently at polynomial cost and are very sparse in the natural orbital basis \cite{khamoshi_efficient_2019}. To see this, first note that for a generic many-body operator $\hat{O}$ we have 
\begin{align}
    \Eagp{\hat{O}} = \Eagp{\hat{O}^{\delta \Omega= 0}}.    
\end{align}
In other words, since AGP is a seniority-zero wavefunction, all many-body density matrices over AGP can be written in terms of $\Eagp{\Pdag{p}...\N{q}...\Pp{r}...}$ where the overlaps associated with particle-number breaking or seniority non-conserving operators are identically zero. Therefore, we can organize all nonzero reduced density matrices in ascending rank as follows 
\begin{subequations}\label{eq:allrdms}
\begin{align}
    Z_{p}^{(1,1)}   &= \Eagp{\N{p}}, \\
    Z_{pq}^{(0,2)}  &= \Eagp{\Pdag{p}\Pp{q}}, \\
    Z_{pq}^{(2,2)}  &= \Eagp{\N{p}\N{q}}, \\
    Z_{pqr}^{(1,3)} &= \Eagp{\Pdag{p}\N{q}\Pp{r}}, \\
    Z_{pqr}^{(3,3)} &= \Eagp{\N{p}\N{q}\N{r}}, \\
    \ldots \nonumber
\end{align}
\end{subequations}
where the first integer in the superscript indicates the number of $\N{p}$ operators in the middle and the second integer is the rank of the density matrix. When indices of an RDM are different, we refer to it as an \textit{irreducible} density matrix. This is because $\N{p} = \N{p}^2/2 = 2\Pdag{p}\Pp{p}$ together with $\Pdag{p}\Pdag{p} = \Pdag{p} \N{p} = 0$ in the seniority-zero space imply that RDMs with repeated indices are either zero or can be written in terms of lower rank density matrices. 

Every matrix element of an AGP RDM, regardless of the rank, can be computed directly at \bigO{M^2} cost using a method based on elementary symmetric polynomials (ESP) \cite{khamoshi_efficient_2019}. Even more efficiently, one can take advantage of the \textit{reconstruction formulae} to express any irreducible RDM as a linear combination of lower rank ones, which can be performed all the way down to occupation numbers and geminal coefficients \cite{khamoshi_efficient_2019}. This allows us to compute any density matrix of rank 2 or higher with \bigO{1} cost per element. Finally, we can take advantage of the equivalence between AGP and the  number-projected BCS wavefunction \cite{ring_nuclear_1980, dukelsky_structure_2016} to extract AGP density matrices via grid integrations of BCS transition density matrices \cite{dutta_construction_2021}, at \bigO{N_\mathrm{grid}} cost per element, where $N_\mathrm{grid}$ is the size of the numerical quadrature for gauge integration \cite{scuseria_projected_2011}. While less efficient than the reconstruction formulae, this final approach can be very efficient when used to contract density matrices with other factorized quantities. In the subsequent sections, we use these methods to compute all density matrices in our calculations.

\section{Jastrow-AGP}\label{sec:JAGP}
\subsection{Analytic properties}\label{subsec:JAGP_analytic}
Let the $k$-body, or rank-$k$, Hilbert-space Jastrow operator \cite{neuscamman_jastrow_2013} be written as
\begin{align}\label{eq:rank_k_jastorw}
    \Jst{k} &= \frac{1}{2^k}\sum_{p_1<...<p_k} \alpha_{p_1...p_k} \N{p_1}\N{p_2}...\N{p_k},
\end{align}
where the sums run over all orbitals, $\alpha$ is a symmetric tensor and is invariant under the exchange of its indices, and ${1}/{2^k}$ is introduced for convenience. Importantly, since $\N{p}^2 = 2 \N{p}$ in the seniority-zero space, we impose that $\alpha$ is zero if any two indices are the same and eliminate these terms from the sum in \Eq{\ref{eq:rank_k_jastorw}}. 

From the $k$-body Jastrow operator we can define the $k$-body Jastrow-AGP (J$_k$AGP) wavefunction as
\begin{equation}
|\mathrm{J_kAGP}\rangle = \mathrm{e}^{\Jst{k}} |\mathrm{AGP}\rangle
\end{equation}
As we prove in \Apdx{\ref{appx:Jn_proof}}, we do not need to include lower body Jastrow operators in this J$_k$AGP wavefunction, as they are contained inside $\Jst{k}$.  Thus, for example, we may write
\begin{equation}\label{eq:J2prime}
\mathrm{e}^{\Jst{1} + \Jst{2}} |\mathrm{AGP}\rangle = \mathrm{e}^{\Jst{2}^\prime} |\mathrm{AGP}\rangle.
\end{equation}
where $\Jst{2}^{\prime}$ is another two-body Jastrow operator.

Slater determinants built from the same orbitals used to define the Hilbert-space Jastrow operators are eigenfunctions of $\Jst{k}$, that is
\begin{subequations}
\label{Eqn:JEigenvalues}
\begin{align}
\Jst{k}|\Phi_\mu\rangle &= J_{k,\mu} |\Phi_\mu\rangle,
\\
J_{k,\mu} &= \frac{1}{2^k} \, \sum_{p_1 < \ldots < p_k} \alpha_{p_1\ldots p_k} N_{p_1,\mu} \ldots N_{p_k,\mu}
\\
 &= \sum_{i_1 < \ldots < i_k} \alpha_{i_1 \ldots i_k},
\end{align}
\end{subequations}
where $|\Phi_\mu\rangle$ is a determinant, $N_{p,\mu}$ is the occupation number of level $p$ in determinant $\mu$, and the indices $i_1 \ldots i_k$ run over the occupied orbitals in $|\Phi_\mu\rangle$. From this, it follows that
\begin{equation}
\mathrm{e}^{\Jst{k}} \sum c_\mu |\Phi_\mu\rangle = \sum \mathrm{e}^{J_{k,\mu}} \, c_\mu \, |\Phi_\mu\rangle.
\label{Eqn:JPsiForm}
\end{equation}
For a complex $\alpha$, the magnitude of the coefficient of a given determinant is modified by the Hermitian part of $\Jst{k}$, i.e. the real part of $\alpha$. The anti-Hermitian part of $\Jst{k}$ (the imaginary part of $\alpha$) only adjusts the phase of each coefficient. It is thus apparent that for an $N$-body initial wavefunction $|\Psi\rangle$ in which all determinants $|\Phi_\mu\rangle$ have non-zero coefficients, $\exp(\Jst{N}) |\Psi\rangle$ can be made an exact eigenstate of a chosen Hamiltonian provided that $\Jst{N}$ is non-Hermitian. 

AGP is, in general, such an initial wavefunction, so $\exp(\Jst{N})$ acting on it can be made exact for the seniority zero space. For any $k \leq N$, it follows from \Eq{\ref{Eqn:JEigenvalues}} and \Eq{\ref{Eqn:JPsiForm}} that
\begin{multline} \label{eq:expJk_agp}
    \mathrm{e}^{\Jst{k}}\rAGP = \sum_{ p_1<...<p_N} \exp\left( \sum_{1\leq j_1 < ... < j_k \leq N} \alpha_{p_{j_1}...p_{j_k}}\right) \\ 
    \eta _{p_1}...\eta _{p_N} \Pdag{p_1}...\Pdag{p_{N}} \rVac.
\end{multline}
In contrast, we can see how the linear wavefunction $\ket{\Psi}= \Jst{k}\rAGP$, which we refer to as J$_k$CI AGP \cite{dutta_geminal_2020}, modifies AGP as follows
\begin{multline}\label{eq:Jk_agp}
    \ket{\Psi} = \sum_{ p_1<...<p_N} \left( \sum_{1\leq j_1 < ... < j_k \leq N} \alpha_{p_{j_1}...p_{j_k}}\right) \eta _{p_1}...\eta _{p_N} \\ 
    \Pdag{p_1}...\Pdag{p_{N}} \rVac.
\end{multline} 
Thus, the way in which the linear and non-linear $\Jst{k}$ correlators factorize the DOCI coefficients can be easily deduced from \Eq{\ref{eq:expJk_agp}} and \Eq{\ref{eq:Jk_agp}}. Together with \Eq{\ref{eq:DOCI}}, we can readily see that
\begin{subequations} \label{eq:doci_approx}
    \begin{gather}
        D_{p_1...p_N}^{\text{(Linear)}} \approx \sum_{1\leq j_1 < ... < j_k \leq N} \alpha_{p_{j_1}...p_{j_k}} \eta _{p_1}...\eta _{p_N}, \label{eq:DOCI_linear} \\ 
        D_{p_1...p_N}^{\text{(Non-linear)}} \approx  \prod_{1\leq j_1 < ... < j_k \leq N} \mathrm{e}^{\alpha_{p_{j_1}...p_{j_k}}} \eta _{p_1}...\eta _{p_N}. \label{eq:DOCI_nonlinear}
    \end{gather}
\end{subequations}
Indeed the two factorizations are different, and \Eq{\ref{eq:DOCI_linear}} can be understood as the linear approximation to \Eq{\ref{eq:DOCI_nonlinear}} up to a constant shift of all amplitudes. An implication can be made, for example, about the sign of each coefficient; if $\alpha$ is real, the DOCI coefficient in Eq. \eqref{eq:DOCI_nonlinear} can be negative only if some $\eta$'s are negative, whereas the same restriction is not implied in \eqref{eq:DOCI_linear}. 

Having shown analytically how $\Jst{k}$ acts on AGP, we concern ourselves with optimizing the energy over JAGP in the subsequent sections. 

\subsection{Energy optimization with JAGP}

Consider $k=1$ for which the J$_1$AGP wavefunction is just an AGP,
\begin{subequations}\label{eq:expJ1_AGP}
\begin{align}
\mathrm{e}^{\Jst{1}} \rAGP
 &= \sum_{p_1 < \ldots < p_N} \mathrm{e}^{\sum_j \alpha_{p_j}} \, \left(\prod_k \eta_{p_k} \, \Pdag{p_k}\right) \rVac
\\
 &= \sum_{p_1 < \ldots < p_N} \left(\prod_k \eta_{p_k} \, \mathrm{e}^{\alpha_{p_k}} \, \Pdag{p_k}\right) \rVac.
\end{align}
\end{subequations}
Evidently, the action of $\Jst{1}$ just changes the geminal coefficient $\eta_p$ to $\eta_p \, \mathrm{e}^{\alpha_p}$. For $k=1$, J$_1$AGP can optimize an AGP, but it cannot correlate it; we need $k>1$ to go beyond an AGP mean-field. Unfortunately, except for $k=1$, the norm and density matrices over $\mathrm{e}^{\Jst{k}}\rAGP$ are not trivial to compute, and aside from quantum Monte Carlo methods, \cite{foulkes_quantum_2001, casula_correlated_2004,neuscamman_communication_2013, neuscamman_improved_2016, wei_reduced_2018, cohen_similarity_2019} we are not aware of an efficient non-stochastic algorithm to compute the overlaps. This for examples makes the variational optimization of JAGP energy
\begin{align}\label{eq:E_varJAGP}
    E_{\text{var-JAGP}} = \frac{\Eagp{\mathrm{e}^{\Jst{}^{\dagger}} \Ham \mathrm{e}^{\Jst{}}}}{\Eagp{\mathrm{e}^{\Jst{}^{\dagger}}\mathrm{e}^{\Jst{}}}},
\end{align}
out of reach for the existing non-stochastic methods. 

However, a case can be made for the similarity transformed Hamiltonian using $\Jst{2}$, that is
\begin{align}
    E_{\text{ST-J$_2$AGP}} = {\Eagp{ \mathrm{e}^{-\Jst{2}} \Ham \mathrm{e}^{\Jst{2}}}}.
\end{align}
The two-body Jastrow correlator has a long history in electronic structure methods and has shown to be capable of recovering a significant portion of the correlation energy. \cite{casula_geminal_2003, neuscamman_communication_2013, genovese_assessing_2019} Previous work in our group has shown that the similarity transformation of fermionic Hamiltonians using a two-body Jastrow operator is summable to all orders and the amplitudes can be solved by left projection via components of $\Jst{2}$ acting on the HF reference \cite{wahlen-strothman_lie_2015, wahlen-strothman_biorthogonal_2016}. Following the same line of reasoning, we extend that formalism to the pairing $su(2)$ algebra \Eq{\ref{eq:CommutationRelations}} and show how to obtain the working equations for AGP in the next section. 

The main idea relies on the fact that the similarity transformation of the Hamiltonian reduces the rank of $\Jst{2}$ by one, thus the remaining one-body Jastrow operator can be subsequently absorbed into the reference. The same idea can be applied to the unitary transformation of the Hamiltonian with an anti-Hermitian
$i\Jst{2}$ operator
\begin{align}\label{eq:E_uJAGP}
    E_{\text{uJ$_2$AGP}} = {\Eagp{ \mathrm{e}^{-i\Jst{2}} \Ham \mathrm{e}^{i\Jst{2}}}},
\end{align}
wherein $\alpha$ is assumed to be real. Based on the argument in the previous section, the Jastrow correlators with purely imaginary amplitudes only contribute to the phase of each Slater determinant in AGP and do not alter their magnitudes. This can also be seen from \Eq{\ref{eq:expJk_agp}} by $\alpha \rightarrow i\alpha$. The phase contributions could be important for Hamiltonians that contain complex terms including those that break time-reversal symmetry. For other Hamiltonians, the mere change in phase is insufficient to make the wavefunction exact.

\subsection{Similarity transformation with JAGP}\label{subsec:STJAGP}
In this section, we extend the results of \Reference{\cite{wahlen-strothman_lie_2015, wahlen-strothman_biorthogonal_2016}} to the AGP wavefunction and our $su(2)$ algebra. For pedagogical reasons, we keep this section self-contained. Suppose one had a generic Jastrow operator $\Jst{}$.  In a manner similar to coupled cluster theory, we intend to solve for the ground state energy by similarity transforming the Hamiltonian,
\begin{subequations}
  \label{eqs:jagp_E}
    \begin{align}
        \simHam &= \mathrm{e}^{-\Jst{}} \Ham \mathrm{e}^{\Jst{}}, \\
        \bar{E} &= \Eagp{\simHam}.
    \end{align}
\end{subequations}
Similarity transformation does not change the spectrum of $\Ham$, but $\simHam$ is no longer Hermitian and $\bar{E}$ is not variational. Therefore we obtain the residual equations by left projecting the Schr\"odinger equation to get
\begin{equation}
\E{\N{p_1} \ldots \N{p_k} \, \simHam} - \bar{E} \, \E{\N{p_1} \ldots \N{p_k}} = 0 \quad \forall_{p_1 < \ldots < p_k}
\label{eqs:res_JAGP}
\end{equation}
where the expectation values are taken with respect to AGP. In traditional coupled cluster theory using particle-hole excitations, one expresses the similarity transformation by the Baker-Campbell-Hausdorff (BCH) expansion whose series naturally truncates at the 4th commutator for a two-body Hamiltonian. In contrast, the transformation using the Jastrow operator does not truncate, but it is summable to all orders for any Hamiltonian.

Our ultimate goal is to use a two- or higher-body Jastrow operators to similarity transform the Hamiltonian and provide correlation. However, it will prove useful to begin by working through the case in which we use a one-body operator, for which the algebra is much simpler. Therefore, consider the $\Jst{1}$ operator
\begin{equation}
\Jst{1} = \frac{1}{2} \, \sum_p \alpha_p \, \N{p}.
\end{equation}
Using \Eq{\ref{eq:CommutationRelations}}, it is easy to show that
\begin{subequations}\label{eqs:J1-ST}
\begin{align}
    \mathrm{e}^{-\Jst{1}} \; \Pdag{p} \; \mathrm{e}^{\Jst{1}} &= \mathrm{e}^{-\alpha_p} \Pdag{p}, \\
    \mathrm{e}^{-\Jst{1}} \; \Pp{p} \; \mathrm{e}^{\Jst{1}} &= \mathrm{e}^{\alpha_p} \Pp{p}, \\
    \mathrm{e}^{-\Jst{1}} \; \N{p} \; \mathrm{e}^{\Jst{1}} &= \N{p}.    
\end{align} 
\end{subequations}
Thus, the similarity-transformed Hamiltonian simply adjusts the coefficients of the $\Pdag{}$ and $\Pp{}$ operators.

\begin{figure*}[t]
    \centering
    \subfloat{{\includegraphics[width=9cm]{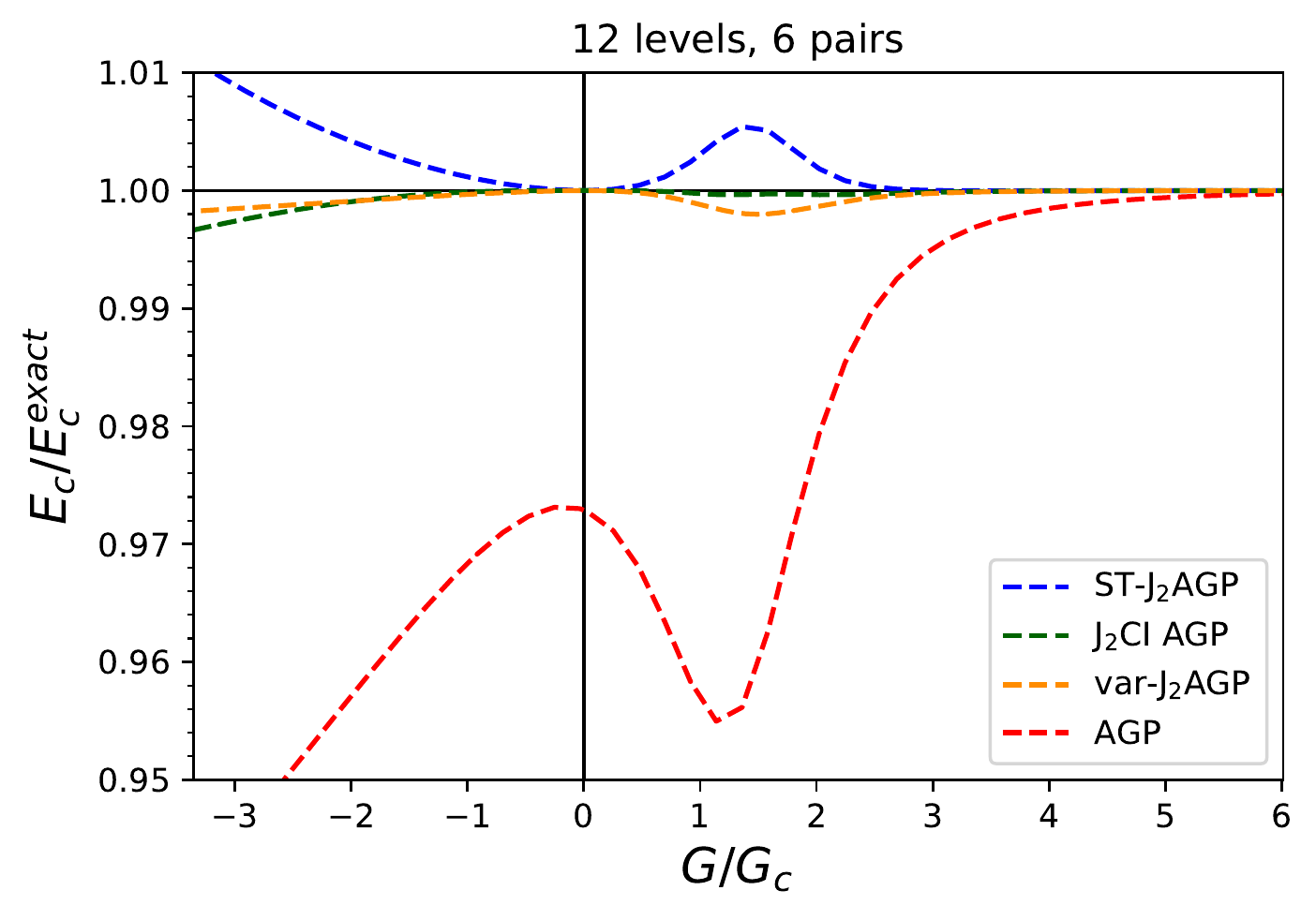}}}
    \subfloat{{\includegraphics[width=9.1cm]{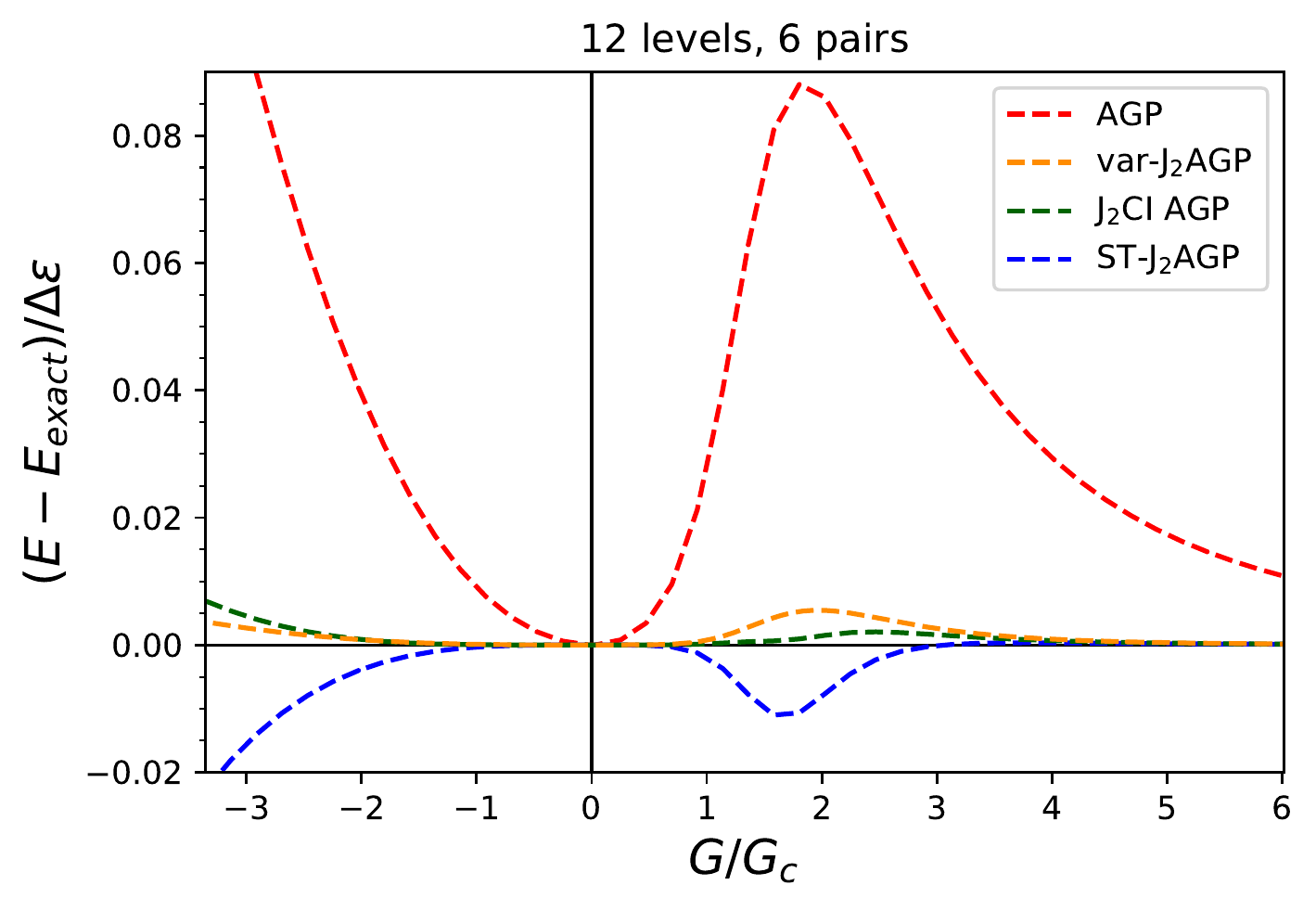}}}
    \caption{Comparison of energy obtained using ST-J$_2$AGP at 12 levels, half-filled with J$_2$-CI and var-J$_2$AGP. The left figure is the correlation energy as the difference with HF and the right is the total energy error.}%
    \label{fig:JAGP_12L6P}
\end{figure*}

Alternatively, we can use \Eq{\ref{eq:expJ1_AGP}} to  absorb $\exp(\Jst{1})$ into AGP. Although we have seen how this can be done, we will extract the same result in an different fashion. Notice by \Eq{\ref{eq:AGP_as_geminal}} that
\begin{align}
    \mathrm{e}^{\Jst{1}} \rAGP = \frac{1}{N!} \left( \mathrm{e}^{\Jst{1}} \GamD \mathrm{e}^{-\Jst{1}} \right)^N \rVac,
\end{align}
where we used the fact that $\exp(\Jst{1})\rVac = \rVac$. From the algebra, \Eq{\ref{eq:CommutationRelations}}, it follows
\begin{subequations}\label{eqs:Gam_ST_J1}
    \begin{align}
        \mathrm{e}^{\Jst{1}} \GamD \mathrm{e}^{-\Jst{1}} &= \sum_p \eta_p \mathrm{e}^{\alpha_p} \Pdag{p}, \\
        \mathrm{e}^{\Jst{1}} \Gam \mathrm{e}^{-\Jst{1}} &= \sum_p \eta_p \mathrm{e}^{-\alpha_p} \Pp{p},
    \end{align} 
\end{subequations}
from which it is clear that $\mathrm{e}^{\Jst{1}}\rAGP$ leads to $\eta_p\rightarrow \eta_p \mathrm{e}^{\alpha_p}$ for all geminal coefficients in the ket, and $\lAGP \mathrm{e}^{-\Jst{1}}$ results in $\eta_p \rightarrow \eta_p \mathrm{e}^{-\alpha_p}$ in the bra.

We now proceed to derive the equations for similarity transformation with $\Jst{2}$. Define $\Jst{1}(p)=\sum_{q}\alpha_{pq} \N{p}/4$ and recall $\alpha_{pp} = 0$, then as in \Eq{\ref{eqs:J1-ST}} we obtain
\begin{subequations}\label{eqs:J2-ST}
    \begin{align}
        \mathrm{e}^{-\Jst{2}} \; \Pdag{p} \; \mathrm{e}^{\Jst{2}} &= \mathrm{e}^{-\Jst{1}(p)} \; \Pdag{p} \; \mathrm{e}^{-\Jst{1}(p)}, \\
        \mathrm{e}^{-\Jst{2}} \; \Pp{p} \; \mathrm{e}^{\Jst{2}} &= \mathrm{e}^{\Jst{1}(p)} \; \Pp{p} \; \mathrm{e}^{\Jst{1}(p)}, \\
        \mathrm{e}^{-\Jst{2}} \; \N{p} \; \mathrm{e}^{\Jst{2}} &= \N{p}    
    \end{align} 
\end{subequations}
where we used $\Commute{\Jst{1}(p)}{\Pdag{p}}=0$ to symmetrically distribute the exponentials in the right hand side of \Eq{\ref{eqs:J2-ST}} on either side of each operator. In fact, using these equations, we can transform any many-body operator of the form $\Pdag{p}...\N{q}...\Pp{r}...$ in a symmetric way. In general, it can be shown that    
\begin{multline} \label{eq:manybody_ST_J2}
    \mathrm{e}^{-\Jst{2}} \; \Pdag{p_1}...\Pdag{p_n}\N{q_1}...\N{q_m}\Pp{r_1}...\Pp{r_n} \; \mathrm{e}^{\Jst{2}} = \\ 
    \mathrm{e}^{\sum_{i=1}^{n} \Jst{1}(r_i)-\Jst{1}(p_i)} \Pdag{p_1}...\Pp{r_n} \mathrm{e}^{\sum_{i=1}^{n}\Jst{1}(r_i)-\Jst{1}(p_i)}.
\end{multline}
To obtain a closed-form expression for the energy and the residual equations, we need to further absorb each $\Jst{1}(p)$ into AGP. Following the same steps as in \Eq{\ref{eqs:Gam_ST_J1}}, we can see that
\begin{multline} \label{eq:Gam_ST_J2}
    \mathrm{e}^{\Jst{1}(q) - \Jst{1}(p)} \GamD \mathrm{e}^{\Jst{1}(p) - \Jst{1}(q)} \\ 
    = \sum_r \left(\eta_r \mathrm{e}^{(\alpha_{qr}-\alpha_{pr})/2} \right) \Pdag{r} = \widetilde{\mathbf{\Gam}}^{\dagger}_{pq}. 
\end{multline}
This together with \Eq{\ref{eq:manybody_ST_J2}} show that we can evaluate the expectation value of any similarity transformed operator of the form $\Pdag{p}...\N{q}...\Pp{r}...$ with $\Jst{2}$ by scaling the geminal coefficients appropriately.  

The same basic steps follow for even higher-body Jastrow operators; the initial similarity-transformation of the Hamiltonian leads to modified matrix elements along with exponentials of Jastrow operators whose ranks are reduced by one. However, only $\Jst{2}$ permits easy evaluation of the remaining expectation values.

For unitary JAGP, as in \Eq{\ref{eq:E_uJAGP}}, one follows the same steps as above but replaces $\alpha \rightarrow i\alpha$. 

\subsection{Application to the pairing Hamiltonian}\label{subsec:STJAGP_app}
We now show how to apply the equations above to similarly transform the pairing  \Eq{\ref{eq:bcs_Hamiltonian}}. It follows form \Eq{\ref{eq:manybody_ST_J2}} that
\begin{multline}
    \mathrm{e}^{-\Jst{2}} \Ham \mathrm{e}^{\Jst{2}} = \sum_p \epsilon_p \N{p} \\ 
    - G \sum_{pq} \left( \mathrm{e}^{\Jst{1}(q)-\Jst{1}(p)} \Pdag{p}\Pp{q} \mathrm{e}^{\Jst{1}(q)-\Jst{1}(p)} \right).
\end{multline}
For every $\Pdag{p}\Pp{q}$, define $|\widetilde{\mathrm{AGP}}_{pq}\rangle =\mathrm{e}^{\Jst{1}(q)-\Jst{1}(p)}\rAGP$, and note that $|\widetilde{\mathrm{AGP}}_{pq}\rangle$ is an AGP whose geminal coefficients have been rescaled by $\tilde{\eta}_r = \eta_r \mathrm{e}^{(\alpha_{qr}-\alpha_{pr})/2}$ for all orbitals $r$ from \Eq{\ref{eq:Gam_ST_J2}}. As such, we can obtain the energy,
\begin{multline}\label{eq:Ebar}
    \bar{E} = \E{\mathrm{e}^{-\Jst{2}} \Ham \mathrm{e}^{\Jst{2}}} = \sum_p \epsilon_p \Eagp{\N{p}}  \\ 
    - G \, \sum_{pq} \langle \widetilde{\mathrm{AGP}}_{pq}|\Pdag{p} \Pp{q}|\widetilde{\mathrm{AGP}}_{pq}\rangle.
\end{multline}
Following similar steps as above, it is straightforward to get closed-form expressions for the residual equations, \Eq{\ref{eqs:res_JAGP}}, as well. We simply use
\begin{multline}
\E{\N{r} \, \N{s} \, \mathrm{e}^{-\Jst{2}} \, \Ham \, \mathrm{e}^{\Jst{2}}} = \sum_p \epsilon_p \, \bra{\mathrm{AGP}}  \N{p} \, \N{r} \, \N{s} \ket{\mathrm{AGP}}
\\
 -G \, \sum_{pq} \langle \widetilde{\mathrm{AGP}}_{pq}|\N{r} \, \N{s} \, \Pdag{p} \, \Pp{q}|\widetilde{\mathrm{AGP}}_{pq}\rangle
\end{multline}
which can then be normal ordered in terms of $\Pdag{p}\ldots\N{q}\ldots\Pp{r}\ldots$.

As outlined in \Sec{\ref{sec:background}}, it is a property of AGP-based RDMs that $Z_p^{(1,1)} = \Eagp{\N{p}}$ is sufficient to construct all higher-body density matrices, where the cost of building $Z^{(1,1)}$ is \bigO{N_\mathrm{grid} M} using grid integration in the number-projected BCS representation of AGP. Of course the same principle applies to RDMs using $|\widetilde{\mathrm{AGP}}_{pq}\rangle$, except that $\widetilde{Z}^{(1,1)}_{r;pq} =\langle \widetilde{\mathrm{AGP}}_{pq}|\N{r}|\widetilde{\mathrm{AGP}}_{pq}\rangle$ needs to be evaluated for every $pq$ in the Hamiltonian, as demonstrated for \Eq{\ref{eq:Ebar}}. The cost of building $\widetilde{Z}^{(1,1)}_{r;pq}$ for all $pqr$ is \bigO{N_\mathrm{grid} M^3} and is the most expensive step in evaluating the ST-J$_2$AGP energy. The cost of evaluating the residuals using the reconstruction formulae is \bigO{M^4}. Without the reconstruction formulae, the cost would be \bigO{M^6}.

In \Fig{\ref{fig:JAGP_12L6P}}, we show the energy error and the fraction of correlation energy captured by ST-J$_2$AGP. All calculations were carried out using the energy optimized AGP for $\Ham$. The correlation energy is computed as the difference with that of HF. As shown in \Fig{\ref{fig:JAGP_12L6P}}, ST-J$_2$AGP over-correlates near the recoupling regime, $G \approx G_c$ and smaller, including $G<0$. In the strongly attractive correlated limit, $G \gg G_c$, ST-J$_2$AGP is highly accurate and becomes exact for sufficiently large $G/G_c$. It is noteworthy, however, that AGP is an eigenstate of the pairing Hamiltonian for $G/G_c \rightarrow \infty$. For comparison purposes, we have plotted the results for variational J$_2$AGP where we carried out the optimization using an in-house DOCI code. We have also included results using a linear wavefunction $\Jst{2} |\mathrm{AGP}\rangle$ in which the energy is determined variationally; this J$_2$CI-AGP has \bigO{M^6} scaling, or \bigO{M^8} without reconstruction formulae.

For larger systems, the over-correlation of ST-J$_2$AGP near the recoupling regime worsens while the linear variational theory remains well-behaved.  Thus, although ST-J$_2$AGP has a lower cost than other AGP-based methods developed recently in our group, neither var-J$_2$AGP nor ST-J$_2$AGP yield as accurate energies for the pairing Hamiltonian.  While correlators build from particle-hole excitations generally benefit from being put in an exponential operator, it is far from clear that the same is true of correlators built from Hilbert space Jastrow operators. Thus in the next section we discuss an exponential ansatz that we build from different generators.

\section{Unitary pair-hopper Correlator} \label{sec:unitaryPH}
An alternative way of creating an exponential ansatz based on two-body, number preserving operators with the generators of the algebra is to use a linear combination of terms like $\Pdag{p}\Pp{q}$. Unlike the Jastrow operators discussed in \Sec{\ref{sec:JAGP}}, a unitary or similarity transformation using $\Pdag{p}\Pp{q}$ operators is not easily summable nor does it truncate naturally the way it does in CC theory. This is a direct consequence of the commutation relations \Eq{\ref{eq:CommutationRelations}}. As a result, some form of approximation must be made. Recently, we proposed a unitary ansatz that we call \textit{unitary pair-hopper} for an implementation on a quantum computer \cite{khamoshi_correlating_2020}. Benchmark calculations against the ground state energy of the pairing-Hamiltonian show highly accurate results. 

In this section, we seek approximation schemes to this model that can be implemented affordably on a classical computer. In particular, we implement two approaches: First, we use the BCH formula and truncate the commutator expansion at some high order. Second, we apply the canonical transformation theory as studied extensively in \Reference{\cite{yanai_canonical_2006, yanai_canonical_2007, neuscamman_quadratic_2009, neuscamman_review_2010, yanai_extended_2012}} to this problem later in this section. 

\begin{figure*}[t]
    \centering
    \subfloat{{\includegraphics[width=9.1cm]{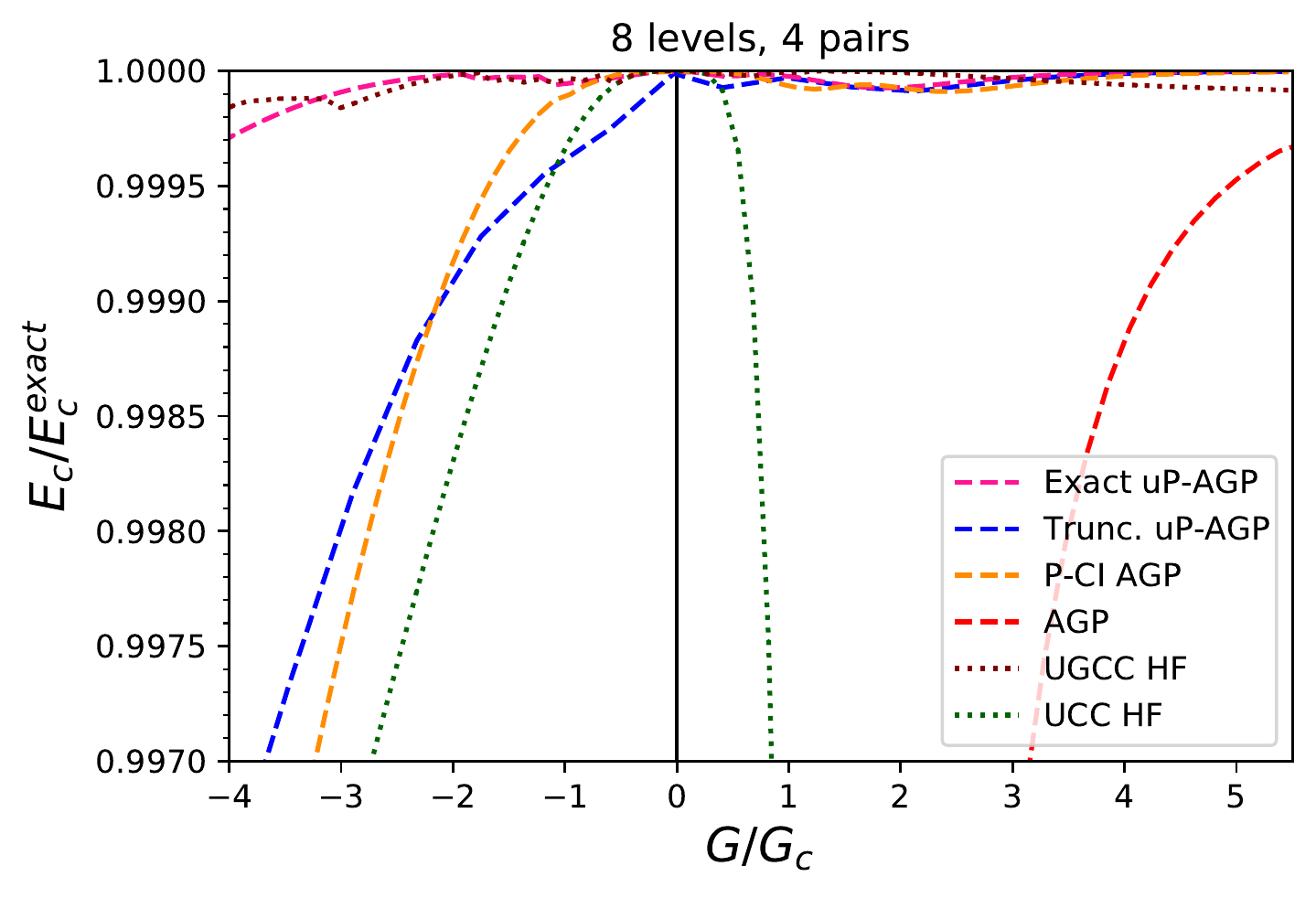}}}
    \subfloat{{\includegraphics[width=9cm]{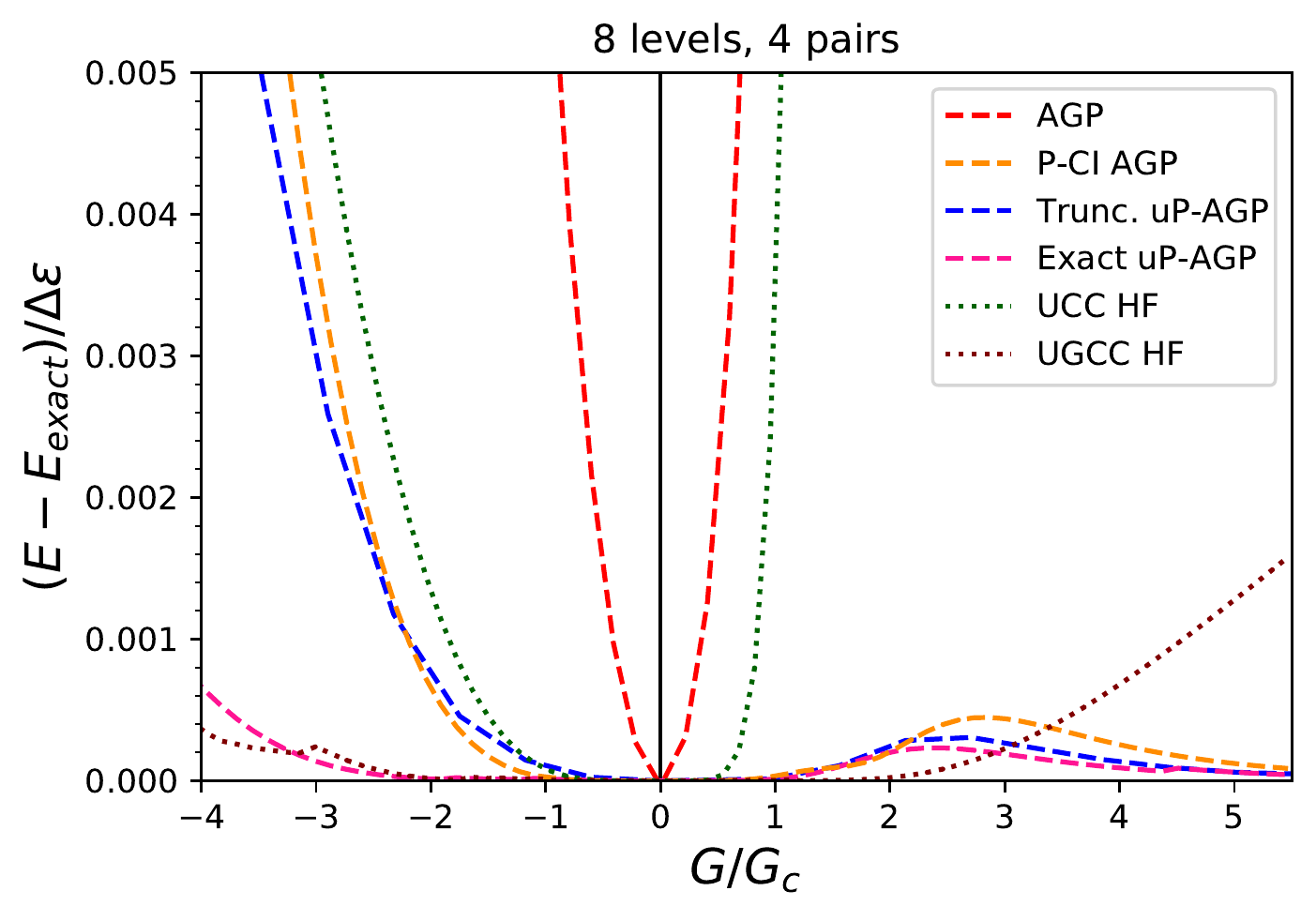}}}
    \caption{Fraction of the correlation energy and the energy error
     of the unitary pair-hopper ansatz, with and without truncation, in comparison with other methods in a small system of 8 levels, half-filled. In addition to AGP based-methods, the plot shows the results for the single-reference unitary coupled cluster on HF. The left figure shows the correlation energy as the difference with HF and the right is the total energy error.}
    \label{fig:pairhopper_8L4P}
\end{figure*}

\subsection{Truncated unitary pair-hopper}\label{subsec:{trunc_PAGP}}
Recall the definition of the anti-Hermitian pair-hopper operator \cite{khamoshi_correlating_2020}
\begin{align}
    \PHoppers = \sum_{p<q} \tau_{pq} \left( \Pdag{p}\Pp{q} - \Pdag{q}\Pp{p} \right),
\end{align}
where $\tau_{pq}$ is the amplitude and is antisymmetric. $\PHoppers$ is derived from 
\begin{align}
    \Killer{pq} - \KillerAd{pq} \propto \Pdag{p}\Pp{q} - \Pdag{q}\Pp{p},
\end{align}
where $\Killer{pq} \rAGP = 0$ is the two-body killer of AGP \cite{henderson_geminal-based_2019}. Its exact expression is
\begin{align}
    \Killer{pq} &= \eta_p^2 \Pdag{p}\Pp{q} + \eta_q^2 \Pdag{q}\Pp{p} + \nonumber \\ 
    &{} \quad \frac{1}{2} \eta_p\eta_q \left( \N{p}\N{q} - \N{p} - \N{q} \right).
\end{align}
Notice that $\mathrm{e}^{\PHoppers}\ket{\text{HF}}$ is equivalent to paired unitary CC with generalized indices (UGCC) \cite{nooijen_can_2000, stein_seniority_2014, lee_generalized_2019}. 

Define the unitary pair-hopper ansatz as $\mathrm{e}^{\PHoppers}\rAGP$ which we denote as uP-AGP. We want to solve for the amplitudes by minimizing the energy, that is
\begin{subequations} \label{eq:minE_ph}
\begin{align} 
    E(\tau) &= \Eagp{\mathrm{e}^{-\PHoppers} \Ham \mathrm{e}^{\PHoppers}}\\
    \tau^{*} &=  \underset{\tau_{pq} \in \mathbb{R}}{\text{argmin}} \; E(\tau).
\end{align}
\end{subequations}
However, since the BCH expansion of the unitary transformed Hamiltonian does not truncate naturally, we choose to truncate the expansion at the highest order that gives us an affordable scaling, i.e. \bigO{M^6}. Again, using the reconstruction formulae, the expectation value of all two-body or higher density matrices can be accessed at \bigO{1} cost per element. This means the asymptotic scaling of the commutator expansion is dominated by the many-body contractions. For example, for a two-body Hamiltonian, the first commutator, $\E{\Commute{\Ham}{\PHoppers}}$ contains three-body contractions; the second commutator, $\E{\Commute{\Commute{\Ham}{\PHoppers}}{\PHoppers}}$ contains four-body contractions, and so on. By truncating the expansion at the 4th commuter, we obtain a scaling of \bigO{M^6} for the energy. In other words, we can state
\begin{multline}
    E(\tau) \approx \underbrace{\E{\Ham}}_{\text{\bigO{M^2}}} 
    + \underbrace{\E{\Commute{\Ham}{\PHoppers}}}_{\text{\bigO{M^3}}}
    + \frac{1}{2!}\underbrace{\E{\Commute{\Commute{\Ham}{\PHoppers}}{\PHoppers}}}_{\text{\bigO{M^4}}} \\
    + \frac{1}{3!}\underbrace{\E{\Commute{\Commute{\Commute{\Ham}{\PHoppers}}{\PHoppers}}{\PHoppers}}}_{\text{\bigO{M^5}}} 
    + \frac{1}{4!}\underbrace{\E{\Commute{\Commute{\Commute{\Commute{\Ham}{\PHoppers}}{\PHoppers}}{\PHoppers}}{\PHoppers}}}_{\text{\bigO{M^6}}},
\end{multline}
such that addition of every term in the expansion multiplies the leading scaling by a factor of $M$. Obtaining the equations above analytically is cumbersome but not prohibitive. We use our home-built algebraic manipulator software, \textit{drudge} \cite{zhao_symbolic_2018}, to generate this and the analytic gradient of the energy. As a result, we can compute both $E(\tau)$ and $\nabla E(\tau)$ at \bigO{M^6} cost albeit with a relatively large prefactor.

We minimize the energy as in \Eq{\ref{eq:minE_ph}} using the limited-memory Broyden-Fletcher-Goldfarb-Shanno (L-BFGS) algorithm \cite{byrd_limited_1995} where we provide the gradients using our analytical expressions. We could alternatively find the solutions to $\nabla E(\tau) = 0$ using non-linear root-finding algorithms, but we find that the former is typically more robust when an accurate initial guess is provided.

Because truncating the commutator expansion yields a non-variational energy expression, we must take care with the minimization of the energy. This becomes particularly important in larger systems for which the numerical algorithms may go below the exact energy and eventually blow up. Fortunately, this failure mode is easy to identify as it is accompanied by extremely large amplitudes in $\tau$. To avoid this problem, we introduce an $L_2$ regularization \cite{nocedal_numerical_2006}
\begin{subequations} \label{eq:regularized_E}
\begin{align}
    \widetilde{E}({\tau}) &= E({\tau}) + \frac{\lambda}{2} \norm{{\tau}}^2 \\
    \implies \nabla\widetilde{E}({\tau}) &= \nabla{E}({\tau}) + \lambda {\tau}
\end{align}
\end{subequations}
where $\lambda>0$ is the regularization parameter and is sufficiently small to allow a robust convergence. In effect, the regularization penalizes the energy optimization in proportion to the magnitude of the amplitudes, thereby preventing it from blowing up. Although this compromises the accuracy of final energy, we gain a smoother convergence. To find $\lambda$, we pick a small value e.g. $10^{-3}$, converge to an optimal solution for $\tau$, and use this solution as the initial guess of another optimization in which a smaller $\lambda$ is used. The first initial guess for $\tau$ can be set to zero. We repeat this process until we can no longer converge or the change in energy is negligible. The convergence threshold of the gradient is set to $10^{-8}$, and we have not observed significant dependence of the final results on the initial guess for $\tau$ or $\lambda$ in our calculations.

In \Fig{\ref{fig:pairhopper_8L4P}} we plot the results for the pairing Hamiltonian in a small system in which we can compare the accuracy of our truncated commutator expansion with that of exact uP-AGP which we obtain from a full configuration interaction (FCI) code. As we can see in the plot, the agreement between the exact and truncated uP-AGP is excellent on the attractive side while at the same time being slightly more accurate than the linear wavefunction $\PHoppers \rAGP$ denoted as P-CI AGP. On the repulsive side, the results for the truncated uP-AGP is comparable to P-CI and becomes more accurate only at sufficiently small $G/G_c$.

\begin{figure*}[t]
    \centering
    \subfloat{{\includegraphics[width=9cm]{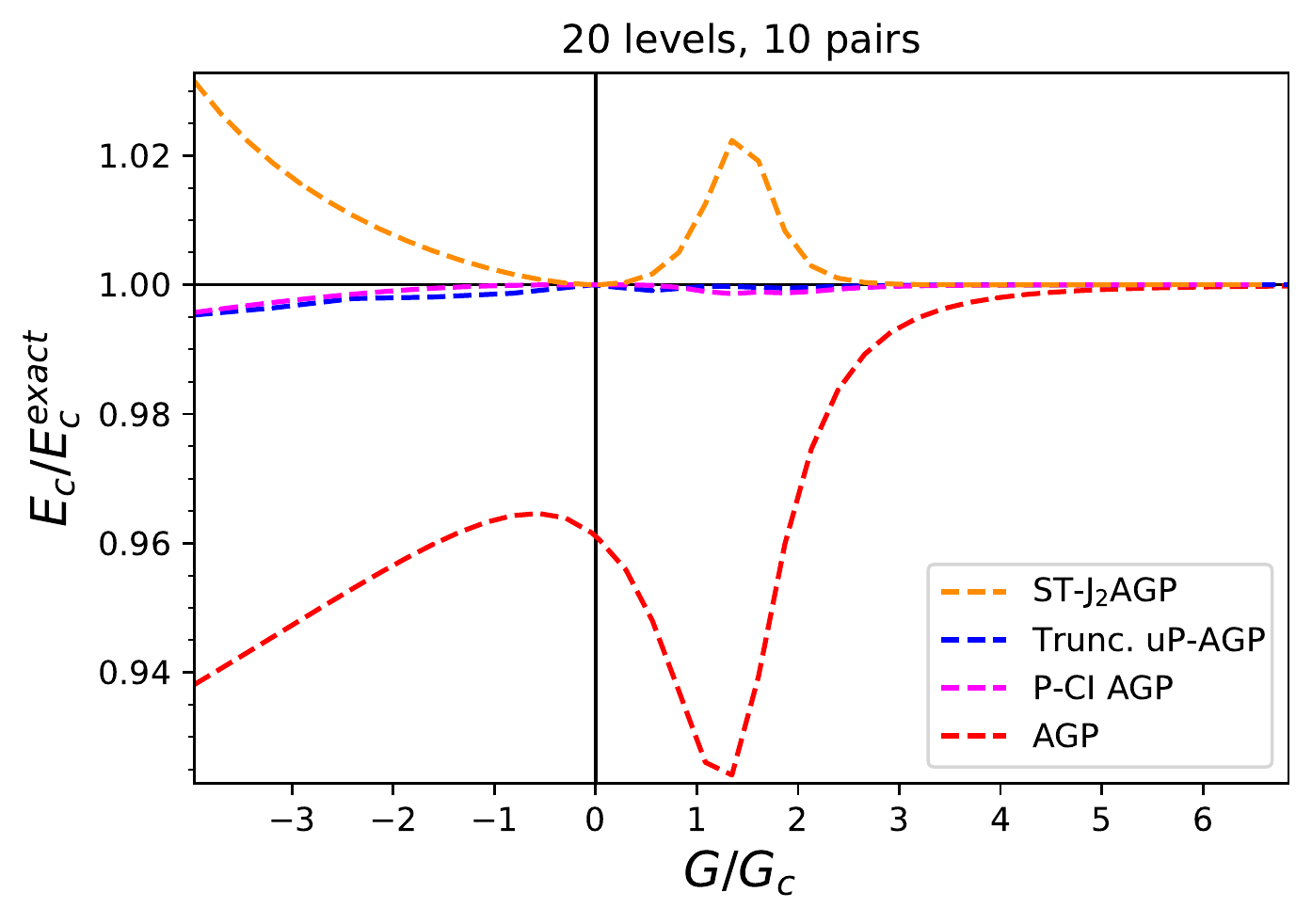}}}
    \subfloat{{\includegraphics[width=9cm]{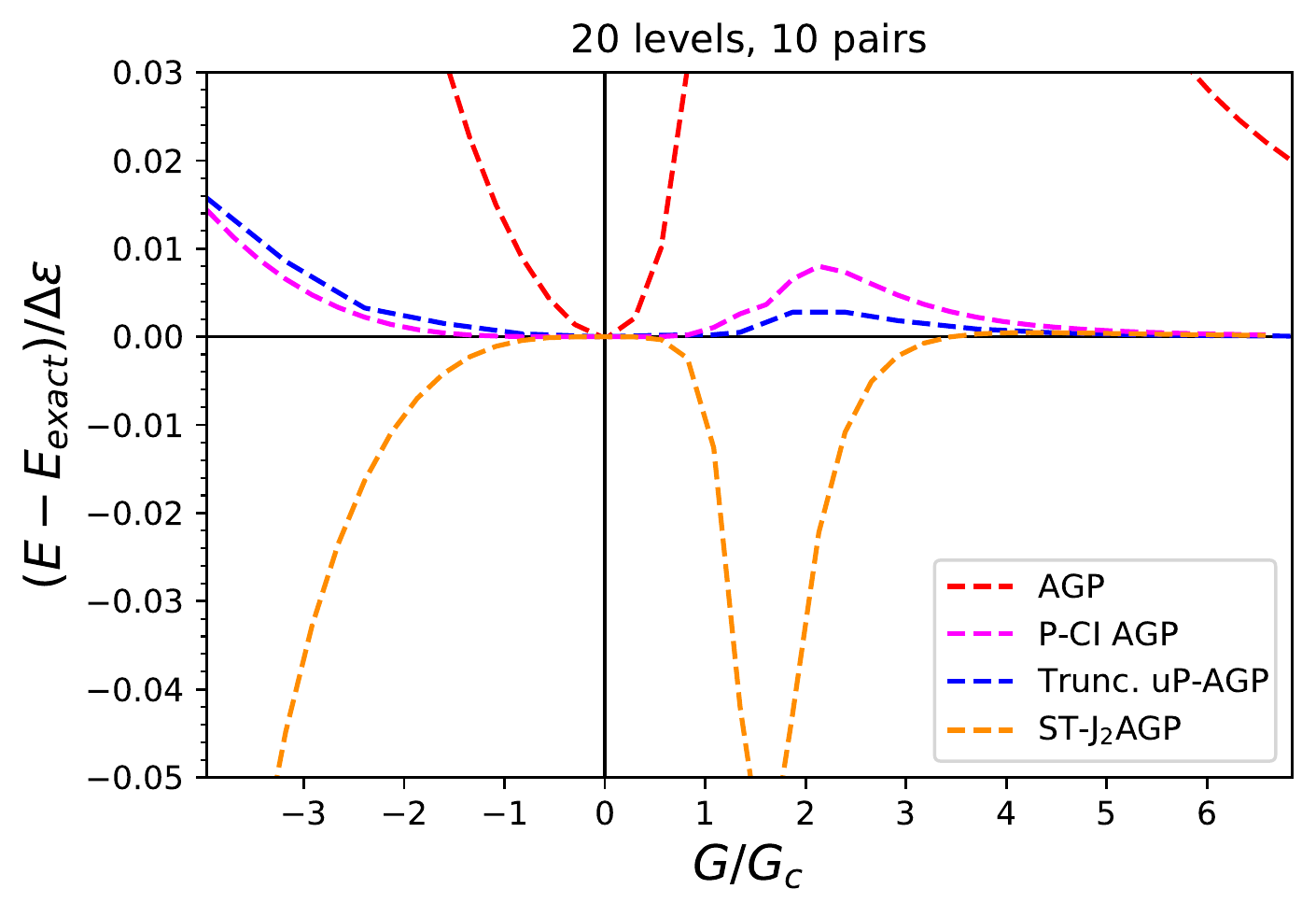}}}
    \caption{Accuracy of the truncated unitary pair-hopper as compared with P-CI AGP and ST-J$_2$AGP at 20 levels, half-filled. The left figure shows the correlation energy as the difference with HF and the right is the total energy error.}
    \label{fig:pairhopper_12L6P}
\end{figure*}

For benchmarking purposes, we also plot HF-based unitary coupled cluster (UCC) \cite{bartlett_alternative_1989, taube_new_2006} as well as that with generalized indices, UGCC, in \Fig{\ref{fig:pairhopper_8L4P}}. The results are obtained from a FCI code. As expected UCC works well in the repulsive regime, but it breaks down as $G$ approaches $G_c$. On the other hand, UGCC is remarkably more accurate than UCC, but it too eventually breaks down for large $G/G_c$. In contrast, our AGP based methods are well behaved in all correlation regimes.

The results for a larger system is show in \Fig{\ref{fig:pairhopper_12L6P}} where we compare the accuracy of the truncated uP-AGP with ST-J$_2$AGP and P-CI. Although all of these methods have the same number of parameters, they clearly differ considerably in their accuracy. The observation that uP-AGP could be more accurate than P-CI in principle is a testament to its non-linear nature. Previously, it was shown that CI methods based on the generators of the algebra, namely J$_2$-CI, P-CI, and K-CI, yield identical results \cite{dutta_geminal_2020}. The other advantage of the non-linear model is that it is agnostic to the linear dependencies of the ansatz. This is true for both ST-J$_2$AGP and the unitary pair-hopper. Although removing the linear dependencies by diagonalizing the metric can yield faster convergence, our numerical results show very little difference in the final energy. 

The downside of the truncated uP-AGP is that it has a relatively large prefactor which is a result of the brute-force expansion of the BCH formula. In the next section we apply canonical transformation theory as another means for approximating the unitary transformed Hamiltonian.

\subsection{AGP-based canonical transformation}\label{subsec:CT-AGP}In canonical transformation (CT) theory, as formulated originally by Yanai and Chan \cite{yanai_canonical_2006, yanai_canonical_2007}, the commutator expansion of a unitary transformed Hamiltonian,  
\begin{align}
    \mathrm{e}^{-\opA} \Ham \mathrm{e}^{\opA} = \Ham + \Commute{\Ham}{\opA} + \frac{1}{2!}\Commute{\Commute{\Ham}{\opA}}{\opA} + \ldots,    
\end{align}
where $\opA$ is an anti-Hermitian operator, is recursively summed by systematically replacing all higher-body operators obtained from each commutator by an approximate, lower-body operator. For example, if $\opA$ and $\Ham$ are each two-body operators, one approximates $\Commute{\Ham}{\opA} \rightarrow \Commute{\Ham}{\opA}_{(1,2)}$ wherein the three-body operator resulting from $\Commute{\Ham}{\opA}$ is approximated in terms of one- and two-body operators. This process yields an effective Hamiltonian, $\simHam \approx \mathrm{e}^{-\opA} \Ham \mathrm{e}^{\opA}$, that, in its simplest form, can be written as

\begin{align}\label{eq:basic_CT_H}
    \simHam = \Ham + \Commute{\Ham}{\opA}_{(1,2)} + \frac{1}{2!} \left[[\Ham, \opA \right]_{(1,2)}, \opA]_{(1,2)} + \ldots,
\end{align}
where we have retained up to two-body terms in the effective Hamiltonian. \Eq{\ref{eq:basic_CT_H}} in particular can be summed recursively by
\begin{subequations}
\begin{align}\label{eq:recursive}
    \simHam^{(n+1)} &= \frac{1}{n} \left[\simHam^{(n)}, \opA \right]_{(1,2)}, \\
    \simHam &= \sum_{n=0} \simHam^{(n)},
\end{align}
\end{subequations}
where $\simHam^{(0)} = \Ham$, and the sum is continued until the change in the parameters of the effective Hamiltonian is sufficiently small. The energy is then defined as $E=\bra{\Psi}\simHam\ket{\Psi}$ where $\ket{\Psi}$ is some mean-field reference. The minimization of the energy can be carried out by solving CC-style amplitude equations $\bra{\Psi} \boldsymbol{\hat{A}}^\dagger_i \simHam \ket{\Psi}$ which can be rearranged to give 
\begin{align}
    R_i = \bra{\Psi} [\simHam, \hat{\opA_i}]  \ket{\Psi} = 0, \quad \forall{\hat{\opA_i} \in \opA},
\end{align}
where $\hat{\opA_i}$ are the operator parts of $\opA$.
\begin{figure*}[t]
    \centering
    \subfloat{{\includegraphics[width=9.1cm]{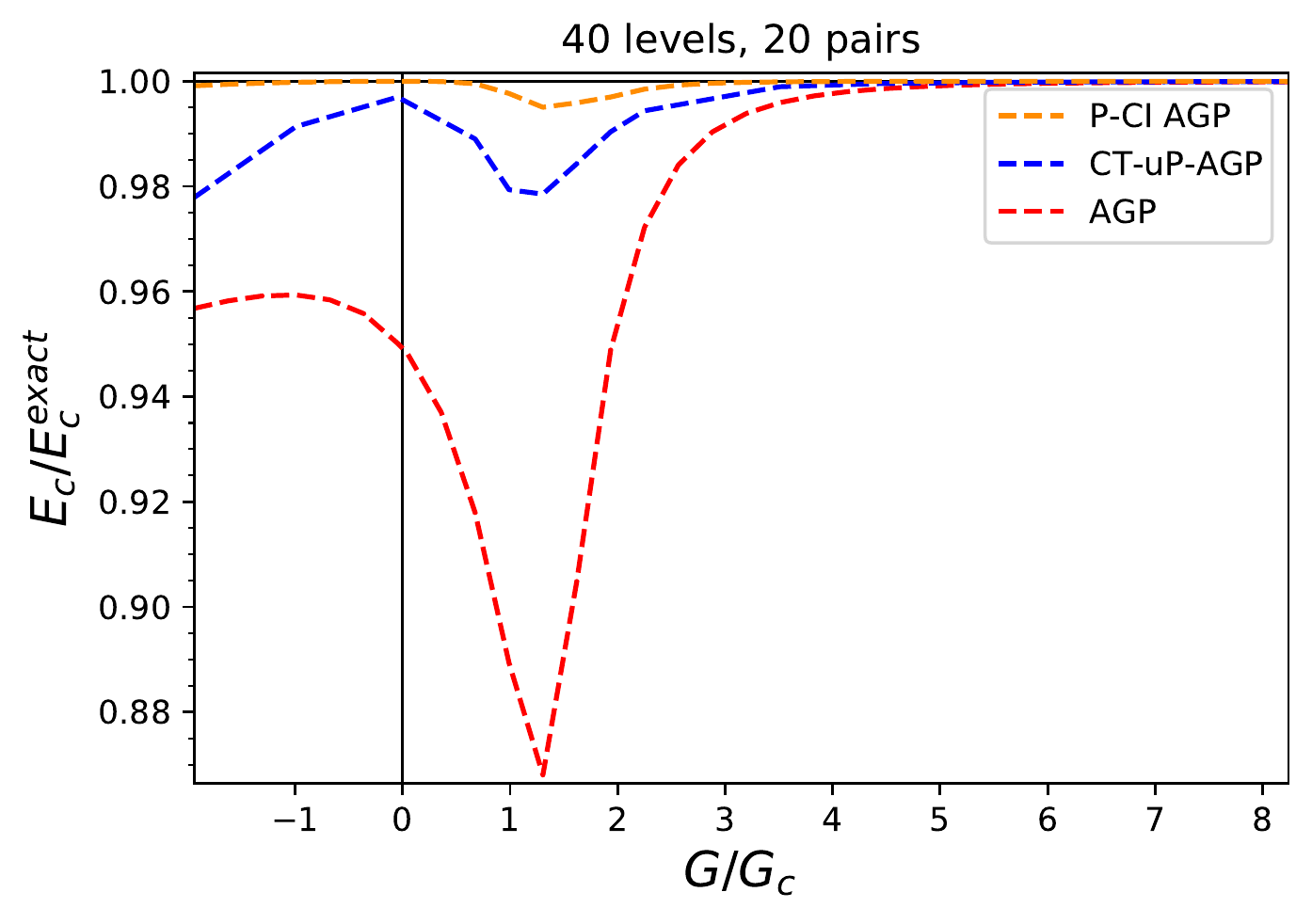}}}
    \subfloat{{\includegraphics[width=9cm]{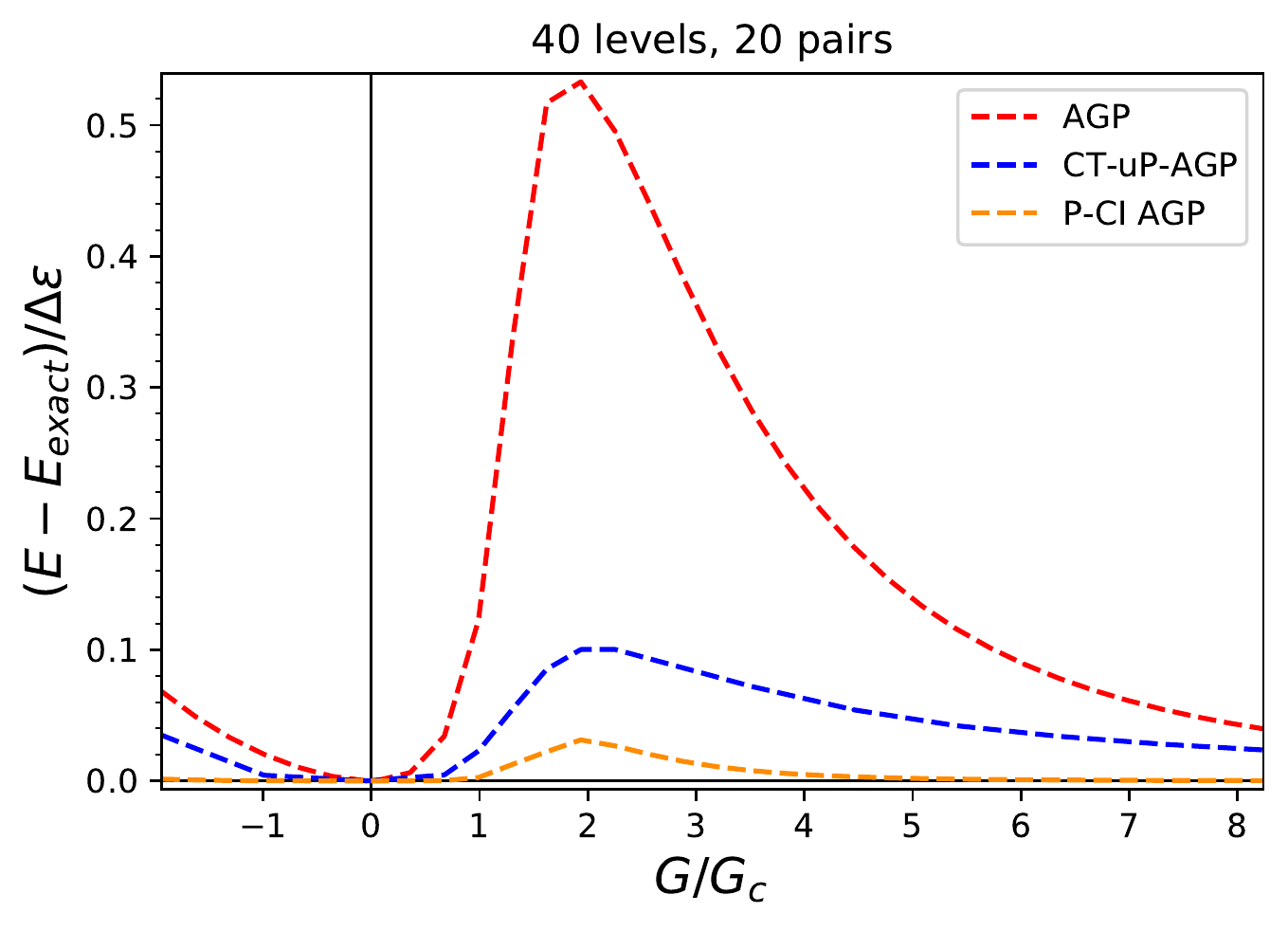}}}
    \caption{Accuracy of canonical transformation with the unitary pair-hoppers, CT-uP-AGP, and that of P-CI AGP. The left figure is the correlation energy as the difference with HF and the right is the total energy error.}
    \label{fig:CTAGP40L20P}
\end{figure*}

In essence, the goal of CT theory is to make the effective Hamiltonian more and more diagonal by successively approximating higher order excitations. It can be viewed as a kind of renormalization group approach and is rooted in an earlier work by White \cite{white_numerical_2002} as well as the flow renormalization approach by Wegner \cite{wegner_flow-equations_1994} and Glazek and Wilson \cite{glazek_perturbative_1994}. In this section, however, we very closely follow the formalism presented in \Reference{\cite{yanai_canonical_2006, yanai_canonical_2007, neuscamman_quadratic_2009, neuscamman_review_2010, yanai_extended_2012}} by Yanai, Chan, and Neuscamman. 

The chief difference between our approach and those of earlier papers is in the operator decomposition or \textit{downfolding} scheme. Instead of invoking the cumulant decomposition of RDMs \cite{kutzelnigg_cumulant_1999} as a means to get lower-body operators, we use AGP and our reconstruction formulae. Just as in cumulant decomposition, the reconstruction formulae are statements about the RDMs, but we can extended them to operators as an approximation. 

For example, consider the three-body irreducible density matrix $Z^{(3,3)}_{pqr}$. From the reconstruction formulae \cite{khamoshi_efficient_2019}, it follows that
\begin{multline}
    \E{\N{p}\N{q}\N{r}} = \\ 
    \Lambda_{qp}\Lambda_{rp} \E{\N{p}} + \Lambda_{pq}\Lambda_{rq} \E{\N{q}} + \Lambda_{pr}\Lambda_{qr} \E{\N{r}},
\end{multline}
where $\Lambda_{qp} = 2\eta_q^2/(\eta_q^2 - \eta_p^2)$. For CT, we propose to approximate $\N{p}\N{q}\N{r}$ by removing the expectation values to get 
\begin{align}
    \N{p}\N{q}\N{r} \rightarrow \Lambda_{qp}\Lambda_{rp} \N{p} + \Lambda_{pq}\Lambda_{rq} \N{q} + \Lambda_{pr}\Lambda_{qr} \N{r}.
\end{align}
Note that since the reconstruction formulae apply only to irreducible RDMs, we limit our downfolding to \textit{irreducible operators}, i.e. $\Pdag{p}\ldots\N{q}\ldots\Pp{r}\ldots$ where all indices are different. Indeed, all reducible operators (RDMs) can be written as a sum of irreducible ones. 

To carry out the CT procedure, we take the unitary pair-hopper $\mathrm{e}^{\PHoppers}$ ansatz to be the generator of our transformation, and we seek an effective Hamiltonian of the form 
\begin{align} \label{eq:general_senZero_H}
    \bar{\Ham} = \sum_{p} h_p \N{p} + \sum_{p \neq q} w_{pq} \N{p}\N{q} + \sum_{pq} v_{pq}\Pdag{p}\Pp{q},
\end{align}
where $w_{pq}$ is symmetric and $v_{pq}$ is Hermitian. Note that \Eq{\ref{eq:general_senZero_H}} is the general form of a two-body Hamiltonian in the seniority-zero space \cite{henderson_pair_2015}. In order to get a relatively more accurate approximation in CT, we would like to delay the downfolding as much possible. To this end, following \Reference{\cite{neuscamman_review_2010}}, we split $\bar{\Ham} = \bar{\Ham}_1 + \bar{\Ham}_2$ such that $\bar{\Ham}_1$ and $\bar{\Ham}_2$ are the one- and two-body parts of the effective Hamiltonian respectively, and the downfolding is delayed until four-body operators appear. As such, one obtains 
\begin{subequations}\label{eq:H_downfold}
\begin{align}\label{eq:CT_H}
    \bar{\Ham_1} &= \Ham_1 + [\Ham_1, \PHoppers] + \frac{1}{2!}[[\Ham_1, \PHoppers], \PHoppers]_{(1,2)} \nonumber \\
    &{} \quad + \frac{1}{3!}[[[\Ham_1, \PHoppers], \PHoppers], \PHoppers]_{(1,2)} + ...  \\
    \bar{\Ham_2} &= \Ham_2 + [\Ham_2, \PHoppers]_{(1,2)} + \frac{1}{2!}[[\Ham_2, \PHoppers], \PHoppers]_{(1,2)}  \nonumber \\ 
    &{} \quad + \frac{1}{3!}[[[\Ham_2, \PHoppers], \PHoppers]_{(1,2)}, \PHoppers]_{(1,2)} + ...
\end{align}
\end{subequations}
Similarly, to get accurate residual equations, we split $R_{pq} = R_{1, pq} + R_{2, pq}$ and evaluate the residuals associated with the one- and two-body parts recursively
\begin{subequations}\label{eq:CT_Res}
\begin{align}
    R_{1, pq} &= \E{[\Ham_1, \PHoppers_{pq}]} + \frac{1}{2!}\E{[[\Ham_1, \PHoppers_{pq}], \PHoppers_{pq}]_{(1,2)}} + \ldots, \\
    R_{2, pq} &= \E{[\Ham_2, \PHoppers_{pq}]_{(1,2)}} + \frac{1}{2!}\E{[[\Ham_2, \PHoppers_{pq}], \PHoppers_{pq}]_{(1,2)}}. + \ldots     
\end{align}
\end{subequations}
In both \Eq{\ref{eq:H_downfold}} and \Eq{\ref{eq:CT_Res}} we retain all two body-terms that naturally appear in the commutator expansion and absorb their coefficients into $w_{pq}$ and $v_{pq}$ in \Eq{\ref{eq:general_senZero_H}}. All higher-body operators are downfolded directly into one-body with the intent of making the Hamiltonian more diagonal. The cost of evaluating the effective Hamiltonian and the residuals is \bigO{M^4}.

Having the effective Hamiltonian and the residual equations, one can solve the amplitudes using a non-linear root-finding algorithm. In so doing, we can provide an approximation to the Jacobian matrix as follows
\begin{align}\label{eq:jacobian}
    J_{pq,rs} &= \E{[[\bar{\Ham}, \PHoppers_{rs}], \PHoppers_{pq}]}.
\end{align}
The cost of building the Jacobian is also \bigO{M^4} by using the reconstruction formula. 

We apply the CT as described above to the pairing Hamiltonian \Eq{\ref{eq:bcs_Hamiltonian}}. As encountered in other implementations of CT, we find that the optimization over the amplitude equations is ill-conditioned \cite{neuscamman_quadratic_2009, neuscamman_review_2010, yanai_extended_2012}. These numerical difficulties are attributed to near-zero eigenvalues of the Jacobian and are said to be similar to the intruder states in the second order perturbation theory \cite{neuscamman_review_2010}. Several techniques have been proposed to mitigate these difficulties \cite{neuscamman_review_2010, yanai_extended_2012}. In our implementation, we find that shifting of the amplitude equations \cite{yanai_extended_2012}
\begin{subequations}
\begin{align}\label{eq:shifted_amps}
    \tilde{R}_{\mu} &= R_{\mu} + \lambda \tau_{\mu} \\
    \implies \tilde{J}_{\mu \nu} &= {J}_{\mu \nu} + \lambda \delta_{\mu \nu},
\end{align}
\end{subequations}
where $\mu, \nu$ are flattened indices, to be the most effective. This is analogous to the regularization introduced in \Eq{\ref{eq:regularized_E}}. The shifting parameter $\lambda>0$ can be chosen \textit{ad hoc} \cite{yanai_extended_2012}. While we may follow the same steps as in the previous section to find optimal values for $\lambda$, for convenience, we picked some sufficiently small values with which the equations can converge reliably.

The results for 40 levels, 20 pairs is shown in \Fig{\ref{fig:CTAGP40L20P}}. We picked $\lambda = 1$ for $G>0$ and a larger $\lambda = 2.5$ for $G<0$. CT with the unitary pair-hopper on AGP (CT-uP-AGP) is considerably less expensive than the truncated uP-AGP as well as the linear P-CI AGP. However, the approximation made to higher order excitations compromises the accuracy of CT-uP-AGP. Systematic improvements can be made to CT-uP-AGP by delaying the downfolding even further, but we leave these improvements for future work. 

\section{Conclusions} \label{sec:conclusions}
We explore several non-linear exponential ans\"atze based on AGP in this paper. First, we study Hilbert-space Jastrow correlators on AGP. We show analytically that while a non-Hermitian JAGP exponential ansatz can in principle be made exact, practical computational limitations with non-stochastic methods compel us to make certain approximations. To this end, we extend the formalism of \Reference{\cite{wahlen-strothman_lie_2015}} to AGP and the $su(2)$ algebra, and show how to similarity transform any given Hamiltonian in the seniority-zero space and solve for the amplitudes in a CC-style manner for AGP. Benchmark calculations of this ST-J$_2$AGP on the ground state of the pairing Hamiltonian show significant improvement on AGP but the resulting energies are less accurate compared to their CI counter-part, J$_2$-CI AGP. The variational ansatz, var-J$_2$AGP, is generally more accurate than the similarity transformed one, but it is out of reach for non-stochastic methods to the best of our knowledge. The cost of ST-J$_2$AGP is \bigO{M^4}.

In the second half of the paper, we sought an approximation to the unitary pair-hopper ansatz. Owing to its non-linear nature, exact variational calculations show uP-AGP can be considerably more accurate than the linear P-CI AGP. However, due to its non truncating commutator expansion, some approximation must be made to implement it efficiently on a classical computer. For an approximation to uP-AGP, we expand $\mathrm{e}^{-\PHoppers} \Ham \mathrm{e}^{\PHoppers}$ using the well-known BCH formula up to the highest affordable computational scaling \bigO{M^6}. Numerical results indicate improvement over P-CI AGP on the attractive regime of the pairing Hamiltonian. 

The downside of truncated uP-AGP is that it has a relatively large prefactor in its asymptotic scaling. To resolve this, we implement a slight variation of canonical transformation theory \cite{neuscamman_review_2010} wherein we carry out the operator decomposition from the reconstruction formulae \cite{khamoshi_efficient_2019} instead of the commonly used cumulant decomposition \cite{kutzelnigg_cumulant_1999}. Our CT-uP-AGP is significantly less expensive than either truncated uP-AGP or P-CI AGP, but the approximation made to the higher order excitations noticeably compromises its accuracy. 

Considering the pros and cons of each method, it is far from clear whether exponential ans\"atze for post-AGP methods are necessarily more advantageous than linear correlators. While there are non-linear ans\"atze that are more accurate than their linear counterparts, practical considerations for implementing them could limit their scope. So far, the linear ans\"atze have shown remarkable accuracy, they are straightforward to implement and seem rather resilient to system size, at least in the pairing Hamiltonian. 

In electronic structure theory, exponential ans\"atze typically enjoy more favorable size-extensivity properties. However, we intentionally avoided discussing size-extensivity in this paper since the interactions in the pairing Hamiltonian are infinite-range. This makes the energy a non-linear function of system size and complicates the discussion of size-extensivity. Nevertheless, it must be noted that Neuscamman has shown \cite{neuscamman_size_2012} variational JAGP is fully size-consistent. Also, earlier work in our group has shown a significant but incomplete restoration of size-consistency  using K-CI and linearized-CC correlators on AGP \cite{henderson_correlating_2020} with the latter exhibiting more favorable size-consistency. A more thorough analysis regarding size-extensivity of post-AGP methods will be reported in due time.

\section{Acknowledgments} 
This work was supported by the U.S. National Science Foundation
under Grant No. CHE-1762320. G.E.S. is a Welch Foundation
Chair (Grant No. c-0036). A.K. is thankful to Yiheng Qiu for useful discussions regarding Jastrow AGP and to Gaurav Harsha for assistance with the drudge software. We thank Rishab Dutta for letting us use his JCI-AGP code. 

\section{Data Availability} 
The data that support the findings of this study are available from the corresponding author upon reasonable request.

\appendix

\section{Proof for J$_k$CI Manifold Containing Lower-Rank JCI Manifolds over AGP} \label{appx:Jn_proof}

The J$_k$CI-AGP manifold contains the J$_{k-1}$CI-AGP and thus all lower-rank JCI-AGP manifolds. This has been observed numerically in \Reference{\cite{dutta_geminal_2020}}. Here we provide a proof for the $k=2$ case, which can be readily generalized to higher ranks. Along the same lines, J$_k$AGP does not need to include lower-rank Jastrow operators, which is shown as a corollary.

For an AGP with $N$ pairs of electrons
\begin{align*}
     &\, \Jst{1} \ket{\mathrm{AGP}}\\
    =&\, \frac{1}{2N} \sum_p \N{p} \Jst{1} \ket{\mathrm{AGP}}\\
    =&\, \frac{1}{4N} \sum_{pq} \alpha_q \N{p} \N{q} \ket{\mathrm{AGP}}\\
    =&\, \frac{1}{4N} \sum_{p < q} \left(\alpha_p + \alpha_q\right) \N{p} \N{q} \ket{\mathrm{AGP}}
    + \frac{1}{4N} \sum_p \alpha_p \N{p}^2 \ket{\mathrm{AGP}}\\
    =&\, \frac{1}{4N} \sum_{p < q} \left(\alpha_p + \alpha_q\right) \N{p} \N{q} \ket{\mathrm{AGP}}
    + \frac{1}{2N} \sum_p \alpha_p \N{p} \ket{\mathrm{AGP}}.
\end{align*}
It follows that
\begin{equation}
    \label{Eqn:J1ssJ2}
    \Jst{1} \ket{\mathrm{AGP}}
    = \frac{1}{4} \sum_{p < q} \frac{\alpha_p + \alpha_q}{N - 1} \N{p} \N{q} \ket{\mathrm{AGP}}.
\end{equation}
Namely, $\Jst{1} \ket{\mathrm{AGP}}$ can be expressed as a linear combination of states in J$_2$CI.

From \Eq{\ref{Eqn:J1ssJ2}}, we can show \Eq{\ref{eq:J2prime}} by defining
\begin{equation}
    \alpha'_{pq} = \frac{\alpha_p + \alpha_q}{N - 1} + \alpha_{pq}
\end{equation}
and
\begin{equation}
    \Jst{2}' = \frac{1}{4} \sum_{p < q} \alpha'_{pq} \N{p} \N{q}.
\end{equation}
Generalization to higher-rank cases is straightforward.

\bibliography{MainBib}

\end{document}